\documentclass[aps,prl,twocolumn, showpacs]{revtex4}

\usepackage{epsfig}
\usepackage{graphicx}

\def\be{\begin{equation}}
\def\ee{\end{equation}}
\def\bea{\begin{eqnarray}}
\def\eea{\end{eqnarray}}

\def\Xint#1{\mathchoice
   {\XXint\displaystyle\textstyle{#1}}%
   {\XXint\textstyle\scriptstyle{#1}}%
   {\XXint\scriptstyle\scriptscriptstyle{#1}}%
   {\XXint\scriptscriptstyle\scriptscriptstyle{#1}}%
   \!\int}
\def\XXint#1#2#3{{\setbox0=\hbox{$#1{#2#3}{\int}$}
     \vcenter{\hbox{$#2#3$}}\kern-.5\wd0}}

\def\dashint{\Xint-}

\begin{document}
\newcount\timehh  \newcount\timemm
\timehh=\time \divide\timehh by 60
\timemm=\time
\count255=\timehh\multiply\count255 by -60 \advance\timemm by \count255

\title{Spin Dynamics and Spin Transport}
\author{Emmanuel I. Rashba \cite{Rashba*} }
\affiliation{Depatment of Physics, MIT, Cambridge, Massachusetts 02139, USA}
\date{\today}


\begin{abstract}
Spin-orbit (SO) interaction critically influences electron spin dynamics and spin transport in bulk semiconductors and semiconductor microstructures. This interaction couples electron spin to dc and ac electric fields. Spin coupling to ac electric fields allows efficient spin manipulating by the electric component of electromagnetic field through the electric dipole spin resonance (EDSR) mechanism. Usually, it is much more efficient than the magnetic manipulation due to a larger coupling constant and the easier access to spins at a nanometer scale. The dependence of the EDSR intensity on the magnetic field direction allows measuring the relative strengths of the competing SO coupling mechanisms in quantum wells. Spin coupling to an in-plane electric field is much stronger than to a perpendicular field. Because electron bands in microstructures are spin split by SO interaction, electron spin is not conserved and spin transport in them is controlled by a number of competing parameters, hence, it is rather nontrivial. The relation between spin transport, spin currents, and spin populations is critically discussed. Importance of transients and sharp gradients for generating spin magnetization by electric fields and for ballistic spin transport is clarified. 
\end{abstract}

\maketitle

\section{INTRODUCTION}

Manipulating electron spins at a given location and transporting electron spins between different locations belong to the central problems of semiconductor spintronics and are of critical importance for quantum computing and information processing \cite{Wolf,Gregg,Zutic,ALS}. Among the different concepts of spin injection and spin manipulation that are discussed in the current literature, the approaches based on SO coupling acquire growing attention.

Spin orbit interaction couples electron spins to the electrical component  \mbox{\boldmath $\tilde{E}$}(t) of electromagnetic field. The Hamiltonian in many cases can be presented in a form
\be
H(\mbox{\boldmath$r$},\mbox{\boldmath$k$},\mbox{\boldmath$\sigma$})=H_{orb}(\mbox{\boldmath$k$},\mbox{\boldmath$\sigma$})+H_Z(\mbox{\boldmath$r$},\mbox{\boldmath$\sigma$}),
\label{eq0}
\ee
where \mbox{\boldmath$r$}, \mbox{\boldmath$k$}, and \mbox{\boldmath$\sigma$} are coordinates, momenta, and Pauli matrices, respectively. Each term of the Hamiltonian including the operators of spatial quantities (\mbox{\boldmath$r$} and/or \mbox{\boldmath$k$}) and Pauli matrices produces SO interaction. In Eq.~(\ref{eq0}), the first term $H_{orb}(\mbox{\boldmath$k$},\mbox{\boldmath$\sigma$})$ symbolizes orbital mechanisms of SO coupling dependent on the electron momentum, while the second term $H_Z(\mbox{\boldmath$r$},\mbox{\boldmath$\sigma$})$ symbolizes the Zeeman energy for a system in an inhomogeneous magnetic field or with a spatially dependent $g$-factor.

In the bulk, orbital mechanisms of SO coupling usually play the major role \cite{R60,RS91} and result in a strong EDSR \cite{MBK,Dob83} dominating over the electron paramagnetic resonance (EPR), the magnetic excitation of spin transitions. However, the spatial dependence of the Zeeman interaction may also play role \cite{RS91,PR65,KRS93}. Remarkably, electrical operation of electron spins in Al$_x$Ga$_{1-x}$As quantum wells has been achieved first through the Zeeman mechanism. Kato {\it et al.} \cite{Kato} took advantage of the anomalously small $g$-factor of the bulk GaAs, $g\approx -0.4$, that allowed them to achieve both the strong anisotropy of the $g$-tensor and its spatial dependence across an inhomogeneous quantum well, ${\hat g}={\hat g}(z)$. Meantime, orbital mechanisms of SO coupling are also greatly enhanced in quantum wells as compared with the bulk material because of the lowering the symmetry that results in developing new SO terms in the Hamiltonian. It is shown in Sec.~II that these mechanisms should result in a strong EDSR, especially in materials with large $g$-factors typical of narrow-gap semiconductors \cite{REprl,REapl}. Extraordinary efficiency of an in-plane field \mbox{\boldmath $\tilde{E}$}(t) is one of the basic conclusions of Sec.~II.

After a number of exciting proposals about using SO coupling for electrical spin injection, it has been already achieved experimentally \cite{Marcus,Rokh}. This success pushes forward the problem of the propagation of inhomogeneous spin populations, say wave packets \cite{Mishch}, in media with a SO split spectrum \cite{Silsbee}. It sounds tempting to apply to this problem a notion of spin currents defined similarly to electric currents. E.g., electrical spin injection and spin polarization produced by SO mechanisms and spin transport in such systems have been discussed recently in terms of spin currents driven by ac \cite{Gov,Malsh} and dc \cite{Mur,Sino} electric fields. However, this appealing approach meets serious problems. First, in media with SO interaction spin currents are not conserved, hence, there is no rigorous definition of them. Second, there is no experimental procedure for measuring them. Third, these currents do not vanish even in thermal equilibrium when there is no physical spin transport. Fourth, as distinct from electric charges and electric currents that appear in Maxwellian equations explicitly, only the spin magnetization can be included into the total magnetization \mbox{\boldmath$M$} in a straightforward way; the place of spin currents in macroscopic physics is still to be unveiled. In addition, the properties of these currents found in Refs.~\onlinecite{Mur} and \onlinecite{Sino} were quite unexpected and inspired an active discussion. Consensus about this problem has not been achieved yet. There are interesting attempts to find the limits within which the concept of spin currents, however not rigorous, can be applied. The different option is to concentrate on the magnetization, the quantity that can be both rigorously defined theoretically and accessible to experimental control. In Sec.~III, some of the related problems are discussed.  

\section{SPIN DYNAMICS IN QUANTUM WELLS}

Quantum wells are usually noncentrosymmetric, and the two-fold spin degeneracy of the electron energy spectrum is lifted by two SO terms
\be
{\hat H}_R=\alpha_R(\sigma_x{\hat k}_y-\sigma_y{\hat k}_x),~
{\hat H}_D=\alpha_D(\sigma_x{\hat k}_x-\sigma_y{\hat k}_y).
\label{eq1}
\ee 
${\hat H}_R$ is due to the structure induced asymmetry (SIA) and is known as Rashba term, ${\hat H}_D$ is due to the bulk induced asymmetry (BIA) and is known as Dresselhaus term, $\sigma_x$ and $\sigma_y$ are Pauli matrices, and ${\hat k}_x$ and ${\hat k}_y$ are the components of the momentum in a magnetic field \mbox{\boldmath$B$} \cite{Winkler}. Because the basic results are similar for ${\hat H}_R$  and ${\hat H}_D$, only equations for ${\hat H}_R$ will be presented in what follows. Intensity of EDSR is controlled by the interaction $e(\mbox{\boldmath${\hat r}$}_{\rm so}\cdot\mbox{\boldmath${\tilde{E}}$}(t))$, where $\mbox{\boldmath${\hat r}$}_{\rm so}$ is a properly defined SO contribution to the coordinate operator. When \mbox{\boldmath$B$} is strong enough, SO coupling can be treated as a perturbation, and, for an electron confined in a parabolic quantum well, the matrix elements of $\mbox{\boldmath${\hat r}$}_{\rm so}$ between spin-up and spin-down states can be found.

In a perpendicular electric field, $\mbox{\boldmath${\tilde{E}}$}\parallel{\hat{\bf z}}$, the matrix element of a spin-flip transition is \cite{REprl}
\be
\langle\uparrow\vert \hat{z}_{\rm so}\vert\downarrow\rangle 
=-~{\alpha_R\over{2\hbar}}~
{{\omega_c\omega_s(\omega_c-\omega_s)\sin 2\theta}\over{\omega_c^2\omega_0^2\cos^2\theta-\omega_s^2(\omega_0^2+\omega_c^2-\omega_s^2)}}~.
\label{eq2}
\ee
Here $\omega_0$ is the parabolic confinement frequency,  $\omega_c=\omega_c(\theta)$ and $\omega_s$ are the cyclotron and spin frequencies, respectively, and $\theta$ is the magnetic-field polar angle. When both SIA and BIA contribute to $\langle\uparrow\vert \hat{z}_{\rm so}\vert\downarrow\rangle$, it acquires an azimuth dependence, and the angular dependence of $\langle\uparrow\vert \hat{z}_{\rm so}\vert\downarrow\rangle$ can be used for measuring the ratio $\alpha_R/\alpha_D$. With $\alpha_R\agt 10^{-9}$ eV cm (as typical of InAs quantum wells), $\langle\uparrow\vert \hat{z}_{\rm so}\vert\downarrow\rangle$ is usually considerably larger than the Compton length, $\lambdabar_C=\hbar/m_0c\approx 4\times10^{-9}$ cm, that plays the role of a characteristic length for EPR. Therefore, EDSR is stronger than EPR.

Nevertheless, the matrix element $\langle\uparrow\vert \hat{z}_{\rm so}\vert\downarrow\rangle$ includes a factor $\omega_s/\omega_0$ that is usually small and reduces the EDSR intensity, $\omega_0^2$ in its denominator indicates that the deviation of the system from the strict 2D limit is the critical condition for EDSR, and it vanishes for $\theta=0$, i.e., EDSR can be observed only in a tilted field \mbox{\boldmath$B$}. Measurements with an in-plane electric field, $\mbox{\boldmath${\tilde{E}}$}\perp{\hat{\bf z}}$, allow one to get rid of all these problems. A matrix element similar to $\langle\uparrow\vert \hat{z}_{\rm so}\vert\downarrow\rangle$ equals \cite{REapl}
\widetext
\be
l_R^\parallel=-~
{{\alpha_R}\over{\hbar(\omega_c^2-\omega_s^2)}}
[(\omega_c\cos\theta+\omega_s)\cos(\varphi-\psi)+i(\omega_c+\omega_s\cos\theta)\sin(\varphi-\psi)],
\label{eq3}
\ee
\endwidetext
\noindent 
where $\varphi$ and $\psi$ are azimuths of \mbox{\boldmath$B$} and \mbox{\boldmath${\tilde{E}}$}, respectively. With $\alpha_R\approx10^{-9}$ eV cm, $m\approx 0.05m_0$, and $B\approx 1$ T, we arrive to $l_R^\parallel$ as large as $\approx10^{-5}$ cm. Then the Rabi frequency $\Omega_R=e{\tilde E}{\it l}/\hbar$, that is the basic figure of merit of the spin operation efficiency, equals $\Omega_R\approx10^{10}$ s$^{-1}$ in a field as small as only about ${\tilde E}\approx 0.6$ V/cm. With such a high efficiency, electrical spin operation seems promising even for Si quantum wells despite of the small $\alpha_R$ values typical of them, $\alpha_R\approx 10^{-12}$ eV cm \cite{WJMR}.

Equations (\ref{eq2}) and (\ref{eq3}) were derived for a strong field \mbox{\boldmath$B$} and a perfect quantum well. For a moderate field \mbox{\boldmath$B$}, spin-flip frequency depends on the Landau quantum number, and a random potential results in an inhomogeneous broadening. Under general conditions, the EDSR band consists of two components with the widths about $\tau_p^{-1}$ and $\tau_s^{-1}$, $\tau_p$ and $\tau_s$ being the momentum and spin relaxation times, respectively \cite{MR}. However, when ${\hat{\mbox{\boldmath$v$}}}_{\rm so}$, the SO part of the velocity operator, does not depend on the momentum \mbox{\boldmath$k$}, the total oscillator strength of the spin-flip transition is contained in the narrow band.  Because ${\hat{\mbox{\boldmath$v$}}}_{\rm so}$ does not depend on \mbox{\boldmath$k$} for both ${\hat H}_R$ and ${\hat H}_D$, a strong dynamic narrowing of the EDSR band is expected.

In conclusion, EDSR seems to be a highly promising tool for the electron-spin manipulation in quantum wells.

\section{SPIN TRANSPORT IN MEDIA WITH SPIN-ORBIT COUPLING:  
SPIN FLUX AND SPIN CURRENTS}

Spin currents are widely discussed and are a subject of excitement and controversy. In what follows, my current understanding of the problem is substantiated and summarized.

The notion of {\it the spin flux}, or, what is the same, of {\it a transport spin current}, has an appealing physical meaning as the propagation of an inhomogeneous spin polarization. However, a rigorous definition of it is lacking. Usually, the definition of the flux density $\mbox{\boldmath$j$}_A(\mbox{\boldmath$r$})$ of a physical quantity $A$ originates from the continuity equation for $A$. The equation $\mbox{\boldmath$j$}_A(\mbox{\boldmath$r$})=\rho_A(\mbox{\boldmath$r$})\mbox{\boldmath$v$}(\mbox{\boldmath$r$})$, where $\rho_A(\mbox{\boldmath$r$})$ is the density of $A$ and $\mbox{\boldmath$v$}(\mbox{\boldmath$r$})$ is the velocity of the flow, holds for the mass flux and electric current due to the mass and charge conservation. Because the mass $m$ of a particle and its electric charge $e$ do not depend on the momentum \mbox{\boldmath$k$}, this definition is tantamount to the integration of the mass and charge currents, $m\mbox{\boldmath$v$}(\mbox{\boldmath$k$})$ and $e\mbox{\boldmath$v$}(\mbox{\boldmath$k$})$, over the \mbox{\boldmath$k$}-space with an appropriate distribution function. However, simple relations of this sort are not universal.

 E.g., because the energy conservation law includes the work performed by the pressure $P(\mbox{\boldmath$r$})$, the energy flow density in an ideal fluid equals $\mbox{\boldmath$j$}_\varepsilon(\mbox{\boldmath$r$})=\rho_w(\mbox{\boldmath$r$})\mbox{\boldmath$v$}(\mbox{\boldmath$r$})$ and includes the enthalpy density $\rho_w(\mbox{\boldmath$r$})=\rho_\varepsilon(\mbox{\boldmath$r$})+P(\mbox{\boldmath$r$})$ rather than the energy density $\rho_\varepsilon(\mbox{\boldmath$r$})$ \cite{FlMech}. 

In media with SO interaction, spin is not conserved. Spin dynamics in the effective magnetic field ${\mbox{\boldmath$B$}}_{\rm eff}(\mbox{\boldmath$k$})$ and, hence, spin nonconservation, is central for the Datta and Das spin-transistor concept \cite{DD90}. Therefore, it is well understood that the definition of the Hermitian operator of {\it the spin-current density}
\be
{\hat{\cal J}}_{ij}(\mbox{\boldmath$k$})=[\sigma_iv_j(\mbox{\boldmath$k$})+v_j(\mbox{\boldmath$k$})\sigma_i]/2,
\label{eq4}
\ee
that is similar to the mass and charge currents and is widely used, lacks the proper justification for systems with SO coupling \cite{Mur2,R03,Burkov,Loss04}; here $i,j$ are Cartesian coordinates. To the best of my knowledge, no experiments for measuring the quantities defined by Eq.~(\ref{eq4}) have been proposed.

Another problem with the spin current is related to its behavior with respect to the time inversion, i.e., the $t\rightarrow -t$ transformation. The mass, charge, and energy densities are even (or real in Wigner's terminology) with respect to the $t$-inversion, while \mbox{\boldmath$v$} is odd (imaginary). Therefore, the corresponding currents are $t$-odd and, after the integration over \mbox{\boldmath$k$}, vanish in thermodynamic equilibrium. On the contrary, Pauli matrices \mbox{\boldmath$\sigma$} are $t$-odd, hence, spin currents are $t$-even and their mean values do not necessarily vanish in equilibrium \cite{R03}. Existence of such currents in noncentrosymmetric systems is in agreement with general principles of statistical mechanics \cite{LP81}. 

These observations indicates existence of a gap between the physical notion of a spin flux ($\equiv$ transport spin current) and the formal definition of spin currents through Eq.~(\ref{eq4}).

\subsection{Momentum Current, Momentum Flux, and Knudsen flux}

In this context, it is instructive to begin with considering first a classical quantity, the momentum current $\Pi_{ii}=\Sigma_{\it l}p_i({\it l})v_i({\it l})$ that also is even with respect to $t$-inversion. Here $\it l$ numerates particles inside a unit volume, and $p_i({\it l})$ and $v_i({\it l})$ designate their momenta and velocities. For an equilibrium gas $\Pi_{ii}=P$, the gas-kinetic pressure that is a thermodynamic rather than transport property. A similar conclusion follows from macroscopic arguments \cite{FlMech}. The mass-flux density $\mbox{\boldmath$j$}_m(\mbox{\boldmath$r$})$ of a streaming fluid equals to its momentum density $\rho(\mbox{\boldmath$r$})\mbox{\boldmath$v$}(\mbox{\boldmath$r$})$, $\rho(\mbox{\boldmath$r$})$ being the fluid velocity. Momentum is conserved in an ideal fluid, and the continuity equation $\partial(\rho v_i)/\partial t=-\partial\Pi_{ij}/\partial x_j$ results in the equation $\Pi_{ij}=P\delta_{ij}+\rho v_iv_j$ for the momentum-flux density tensor $\Pi_{ij}(\mbox{\boldmath$r$})$; in equilibrium, its diagonal components reduce to the momentum current $P$ found above. The nondiagonal part ($i\neq j$) of the momentum-flux density $\rho(\mbox{\boldmath$r$}) v_i(\mbox{\boldmath$r$})v_j(\mbox{\boldmath$r$})$, describing the momentum transport, can be found from $\Pi_{ij}$ directly; it vanishes in equilibrium. However, splitting $\Pi_{ii}(\mbox{\boldmath$r$},t)$ into the pressure $P$ and the momentum flux requires solving the set of the fluid-mechanics equations and cannot be performed in a general form. 

Remarkably, for a Knudsen gas flow across a narrow opening, the flux can be easily related to $P$. Indeed, the momentum-flux density equals to the  integral of the momentum-current density $\Pi_{ii}(\mbox{\boldmath$k$})$ over a single hemisphere, hence, the Knudsen momentum flux equals $\Pi_{ii}/2=P/2$, a half of the momentum current. For the Knudsen flux, {\it a large gradient} of the concentration is critical.

\subsection{Spin Currents in Thermodynamic Equilibrium}

The similarity between ${\hat{\cal J}}_{ij}(\mbox{\boldmath$k$})$ and $\Pi_{ij}(\mbox{\boldmath$k$})$ discussed above suggests existence of nonvanishing equilibrium spin currents in noncentrosymmetric systems; they will be termed as {\it background spin currents} in what follows \cite{R03}. For the Hamiltonian ${\hat H}_R$, a calculation renders
\be
{\cal J}_{xx}={\cal J}_{yy}=0,~{\cal J}_{xy}=-{\cal J}_{yx}\equiv{\cal J}_R=m^2\alpha_R^3/3\pi\hbar^5,
\label{eq5}
\ee
whenever the electrochemical potential $\mu>0$. Spin currents carried by separate electrons are linear in $\alpha_R$. The current ${\cal J}_R$ is proportional to $\alpha_R^3$ because of the partial cancelation of the contributions from two spectrum branches, $\varepsilon_\lambda(\mbox{\boldmath$k$})=\hbar^2k^2/2m+\lambda\alpha_Rk$, $\lambda=\pm 1$. When $\mu\gg\varepsilon_\alpha$, where $\varepsilon_\alpha=m\alpha_R^2/\hbar^2$ is a characteristic SO energy, the difference is relatively small. Nevertheless, it persists under the conditions of thermal equilibrium when there is no spin transport. The nonvanishing equilibrium spin current should be taken as a warning indicating problems inherent in the spin current concept.

The real antisymmetric pseudotensor ${\cal J}_{ij}$ is equivalent to a real vector $\mbox{\boldmath$P$}\parallel{\hat{\bf z}}$. This relation implies the existence of some connection between the background spin currents and the SO contribution to the polarization of the 2D electron gas by the electric field $\mbox{\boldmath$E$}_\perp$ producing the SIA. If to define the polarization $P$ as $P=-d{\cal E}_R/dE_\perp$, where ${\cal E}_R$ is the SO energy of the 2D electron gas, then
\be
P=(4m/\hbar)~{\cal J}_R~ d\alpha_R/dE_\perp.
\label{eq6}
\ee
Therefore, spin current ${\cal J}_R$ is directly related to the SO contribution $P$ to the electric dipole moment. Here $E_\perp$ includes not only the field inside the confinement layer, but also the change in the band offsets that strongly influences $\alpha_R$ \cite{ZawPf}. 

Background spin currents are highly sensitive to the symmetry of the system. Numerical calculations suggest that $\alpha_R$ can show an essential $k$ dependence. Accepting $\alpha_R(k^2)=\alpha_0+\alpha_1(k^2)$ with $\alpha_1(k^2)\ll\alpha_0$, one comes to
\be
{\cal J}_{xy}={{m/2\pi\hbar^3}\over{\sqrt{1+2\mu\hbar^2/m\alpha_0^2}}}
\left(\mu+{{m\alpha_0^2}\over{\hbar^2}}\right)
[\alpha_1(k_-^2)-\alpha_1(k_+^2)],
\label{eq6a}
\ee
where $k_+$ and $k_-$ are Fermi momenta for the upper and lower spectrum branches, respectively. When writing Eq.~(\ref{eq6a}), the contribution proportional to $[\alpha_1(k_-^2)+\alpha_1(k_+^2)]/2$ has been subtracted; it can be found by differentiating Eq.~(\ref{eq5}) over $\alpha_R$. For a small $\alpha_0$, when $m\alpha_0^2/\hbar^2\ll\mu\equiv\hbar^2k_F^2/2m$, Eq.~(\ref{eq6a}) simplifies
\be
{\cal J}_{xy}=(mk_F|\alpha_0|/4\pi\hbar^3)[\alpha_1(k_-^2)-\alpha_1(k_+^2)].
\label{eq6b}
\ee
As distinct from Eq.~(\ref{eq5}), this expression is quadratic rather than cubic in SO coupling. This is a result of the symmetry breaking. Indeed, the Hamiltonian ${\hat H}_R$ with $\alpha_R$=const possesses a hidden symmetry that manifests itself in the fact that the square of the velocity operator $\mbox{\boldmath${\hat v}$}=\hbar^{-1}\partial{\hat H}_R/\partial\mbox{\boldmath$k$}$ is related to ${\hat H}_R$ as ${\mbox{\boldmath${{\hat v}}$}}^2=2{\hat H}_R/m+2\alpha_R^2/\hbar^2$. As a result, its eigenvalues $v^2=2\varepsilon/m+2\alpha_R^2/\hbar^2$ depend only on the energy $\varepsilon$ and do not depend on the branch index $\lambda$. A similar relation holds for the semiclassical velocity, $v_{\rm sc}^2=2\varepsilon/m+\alpha_R^2/\hbar^2$. Both equations result in $\lambda$-independent Fermi velocities.

 The Hamiltonian ${\hat H}_R$ is typical of (0,0,1) 2D layers of cubic crystals. For a 2D layer of an uniaxial crystal like CdS with an in-plane  pyroelectric axis ${\hat{\bf c}}\parallel{\hat{\bf y}}$, a new SO term 
\be
H'_{\rm so}=\alpha_z\sigma_zk_x 
\label{eq6c}
\ee
develops in the Hamiltonian. When $\alpha_z\ll\alpha_R$, spin current produced by this term equals \cite{R04}
\be
{\cal J}_{zx}=-k_F^2\alpha_z/2\pi\hbar.
\label{eq7}
\ee 
Therefore, in this case the background spin current ${\cal J}_{zx}$ is linear in the SO coupling constant $\alpha_z$. Background spin currents are confined inside a sample. In a texture consisting of zinc blende and wurtzite crystallites, the current ${\cal J}_{zx}$ is confined inside the latter ones. In equilibrium, the boundaries adjust to ensure this confinement.

It is seen from Eqs.~(\ref{eq5}) -- (\ref{eq7}) that, as distinct from $\Pi_{ij}$, the nondiagonal components of ${\cal J}_{ij}$ do not vanish even in equilibrium, and the lower is the symmetry the larger the background spin currents are. In a different way, equilibrium spin currents were found by Pareek \cite{Pareek}. Similar problems arise in the theory of magnetic materials in inhomogeneous magnetic fields \cite{ML,SKK} what is quite natural because the Hamiltonian $H_Z(\mbox{\boldmath$r$},\mbox{\boldmath$\sigma$})$ of Eq.~(\ref{eq0}) represents one of SO coupling mechanisms.

\subsection{Response to a Time-Dependent Electric Field}

The paper by Sinova {\it et al.} \cite{Sino} has attracted recently an active interest to the spin current ${\cal J}_{zx}$ driven by a homogeneous electric field $\mbox{\boldmath$E$}=E_y{\hat{\bf y}}$. The operators entering the corresponding Kubo formula are
\be
{\hat v}_y=\hbar k_y/m+\alpha_R\sigma_x/\hbar,~~{\hat{\cal J}}_{zx}=\hbar k_x\sigma_z/m.
\label{eq7a}
\ee
Because in the basis of the eigenspinors of the Hamiltonian ${\hat H}_R$ (the chiral basis) the operator ${\hat{\cal J}}_{zx}$ has only interbranch matrix elements, the spin conductivity $\Sigma_{zxy}={\cal J}_{zx}/E_y$ of a perfect crystal originates from the interbranch electronic transitions inside the ring in the \mbox{\boldmath$k$}-space restricted by the Fermi momenta $k_\pm$, Fig.~1.  The real part of it,
\be
\Sigma'_{zxy}(\omega)={{e\alpha_R}\over{2\pi\hbar m}}\dashint_{k_+}^{k_-}
{{k^2~dk}\over{(2\alpha_Rk/\hbar)^2-\omega^2}}~,
\label{eq8}
\ee
in the $\omega\rightarrow0$ limit results in $\Sigma'_{zxy}(\omega=0)=e/4\pi\hbar$ in agreement with Ref.~\onlinecite{Sino}. Remarkably, $\Sigma'_{zxy}(\omega)$ is related to the SO contribution, $\epsilon'_{\rm so}(\omega)$, to the real part of the dielectric function $\epsilon(\omega)$ by a simple equation
\be
\Sigma'_{zxy}(\omega)=(\hbar^3\omega^2/8\pi em\alpha_R^2)~\epsilon'_{\rm so}(\omega).
\label{eq9}
\ee
Hence, for the Hamiltonian ${\hat H}_R$ the spectrum-specific frequency dependence cancels from the ratio $\Sigma'_{zxy}(\omega)/\epsilon'_{\rm so}(\omega)$. Eqs.~(\ref{eq6}) and (\ref{eq9}) indicate existence of an algebraic connection between spin currents and the SO contribution to dielectric polarization. These quantities should be related in some way because in Maxwellian electrodynamics the response of any homogeneous medium to a homogeneous electric field is described by the dielectric function. 

Deriving $\Sigma'_{zxy}(\omega=0)$ in this way from the Kubo formula of Eq.~(\ref{eq8}) is equivalent to the effect of the SO term $H'_{\rm so}$ with $\alpha_z=-eE_y/3k_F^2$. The same equation can be derived by using the perturbation theory in the operator $V=-eE_yy=ieE_y\partial/\partial k_y$ in the $2\times2$ spinor space. It results in a new equilibrium state in the basis of new eigenspinors. From this standpoint, $\Sigma'_{zxy}(\omega=0)$ of Eq.~(\ref{eq9}) is a thermodynamic rather than a kinetic parameter and is similar to the pressure $P$ of Sec.~III.A. {\it Such static spin currents are dissipationless because they do not transport spins}. This conclusion suggests that $\Sigma'_{zxy}(\omega=0)$ calculated by using only interbranch matrix elements should be nearly insensitive to impurity scattering. Of course, one has to bear in mind that the similarity between the two above systems is not complete. In the system by Sinova {\it et al.} \cite{Sino} there is an external homogeneous electric field $E_y$, while in a conducting pyroelectric the macroscopic field is screened and $E_y$ should be understood as an equivalent microscopic field changing the electron energy spectrum.

 \centerline{\epsfig{file=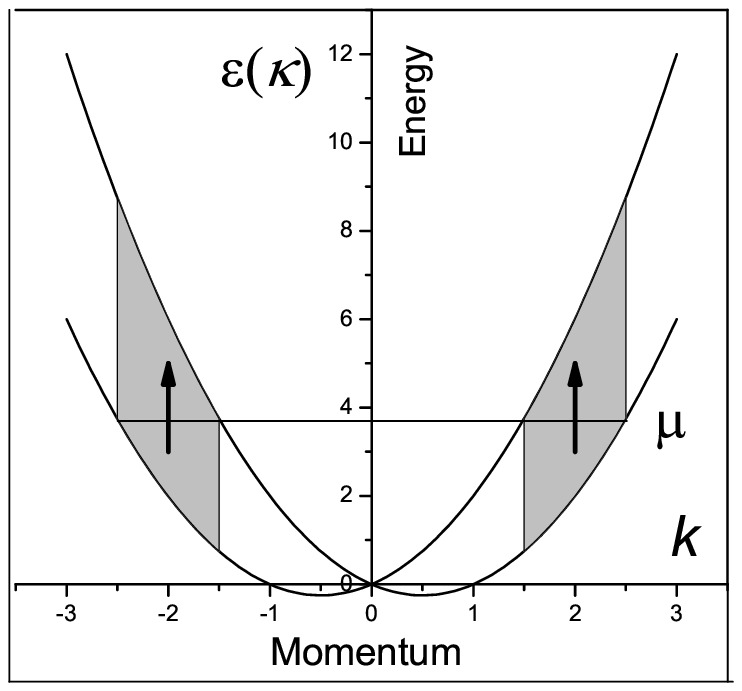,width=8cm}}
 \vskip 0.05 in
 \vbox{\baselineskip0.4 pt
 {\small {\bf Fig.~1} Electronic transitions contributing to spin
currents are shown by arrows inside the gray strips. $\mu$ is the
electrochemical potential, $k_+$ and $k_-$ are the Fermi momenta for
the upper and lower spectrum branches, respectively; $T=0$. Minima of
the spectrum $\varepsilon(k)$ are achieved at the points $\pm
k_\alpha$, where $k_\alpha=m\alpha_R/\hbar^2$.}}

The problem is that the singular intrabranch contributions to $\mbox{\boldmath$q$}=0$ responses (whose explicit form is sometimes elusive) nevertheless exist and manifest itself when the momentum conservation is broken by impurities. E.g., the low-frequency contribution of the Hamiltonian ${\hat H}_R$ to the imaginary part of the dielectric function, 
\be
\epsilon''_{\rm sing}(\omega)=(\pi{\bar\omega}^2_p/2\omega)\delta(\omega),
\label{eq9a}
\ee
comes through the renormalization ${\bar\omega}^2_p=\omega^2_p[1-(m\alpha_R/\hbar^2)/2\pi n]$ of the ``plasma frequency" $\omega_p^2=4\pi e^2n/m$, $n$ being the 2D electron concentration \cite{R03}. This contribution manifests itself in magnetic resonances \cite{R60,RS91}.

\subsection{Generating Spin Fluxes and Spin Polarization}

When a system is out of equilibrium, the states with non-zero spin currents have potentialities for generating spin fluxes. A nonequilibrium state prepared by an electric-field pulse $\mbox{\boldmath$E$}(\mbox{\boldmath$r$},t)$ inside a narrow spot can serve as an example. If the boundaries of the spot are narrow as compared with the electron mean free path $v_F\tau_p$, $v_F=\hbar k_F/m$ being the Fermi velocity, the density of the Knudsen-type spin flux across the boundary can be as high as about $\Sigma'_{zxy}(\omega=0)E/2$, cf. Sec.~III.A. Apparently, spin fluxes observed by Stevens {\it et al.} \cite{Stevens} in their optical experiment were of this sort. 
Strong nonequilibrium near the edge of the spot is critical for generating spin fluxes across it. Knudsen spin fluxes across an abrupt boundary of zinc blende and wurtzite crystallites (Sec.~ III.B), when the pyroelectric field changes faster than the boundary equilibrium establishes, can serve as an  example. Similar spin fluxes have been discussed by Mal'shukov {\it et al.} from a different standpoint \cite{Malsh}.

 Electric field \mbox{\boldmath$E$} is $t$-even while magnetization \mbox{\boldmath$M$} is $t$-odd. Therefore, {\it for an electric field to produce a magnetization} (a magneto-electric effect), $t$-{\it symmetry should be broken} by a proper magnetic structure \cite{Dzy}, electron scattering \cite{IvchP,Levit}, finite frequency, circular polarization, etc. E.g., a dissipationless spin current ${\cal J}_{zx}$ driven by the field $\mbox{\boldmath$E$}=E_y{\hat{\bf y}}$ is not accompanied by any magnetization, however, a magnetization $\mbox{\boldmath$M$}\parallel{\hat{\bf x}}$ proportional to $\tau_p$ does develop in this geometry \cite{Edel}. In microstructures with a Sharvin-type resistance \cite{Sharvin}, the resistance quantum $h/e^2$, $h=2\pi\hbar$, plays the role of the time-symmetry breaking parameter. E.g., the resonant-tunneling-diode spin-filters proposed by Voskoboynikov {\it et al.} \cite{Vosk} are controlled by $\tau_p$ while their modification proposed by Koga {\it et al.} \cite{Koga} might be controlled by $h/e^2$.
 
When $\tau_p\ll\hbar/\alpha_Rk_F$, the magnetization is long living due to the Dyakonov - Perel process \cite{DP}, EDSR line is dynamically narrowed \cite{MR}, quasiequilibrium in the orbital degrees of freedom is established \cite{MR,DP}, and propagation of the magnetization $\mbox{\boldmath$M$}(\mbox{\boldmath$r$},t)$ is controlled by diffusion, drift, and appropriate SO corrections. With increasing $\tau_p$, different scales related to SO coupling come into the game. Spin fluxes not accompanied by magnetization have short decay times about $\tau_p$.

\subsection{Response to a Space- and Time-Dependent Electric Field}

Therefore, transients and spatially inhomogeneous spin populations are critical for spintronics, and spin polarizations are the principal measurables, i.e., quantities accessible to experimental control. 

Spin transport has two major aspects: generation of nonequilibrium spins and their relaxation. There is an extensive literature on the mechanisms of spin relaxation. In this section, I find the spatial and temporal scales related to spin production in media with SO coupling. It can be done by calculating the response of free electrons to a field $\mbox{\boldmath${\tilde{E}}$}\exp\{i[(\mbox{\boldmath$q$}\cdot\mbox{\boldmath$r$})-\omega t]\}$. The static $M_x$-magnetization driven by a field $\mbox{\boldmath$E$}=E_y{\hat{\bf y}}$ \cite{Edel} has been recently observed \cite{Aw04,Gan04}.

The in-phase spin polarization in the frequency range of the interbranch transitions comes from the $\delta$-function parts of Green functions. For $\mbox{\boldmath$q$}=q{\hat{\bf x}}$ and $\mbox{\boldmath$\tilde{E}$}={\tilde E}_y{\hat{\bf y}}$, the mean value of the Pauli matrix $\sigma_x$ reads
\bea
\sigma_x(q,\omega)&=&{{e{\tilde E}_y}\over{16\pi\alpha_R}}I(q,\omega),\nonumber\\
~I(q,\omega)&=&\int_\Phi{{\cos^2\phi~d\phi}\over{(1+q\cos\phi/2k_\alpha)^2}}~.
\label{eq10}
\eea
Here $k_\alpha=m\alpha_R/\hbar^2$ is the shift of the minima of the $\varepsilon(k)$ curves of Fig.~1 from the origin. The integration over the angle $\phi$, the azymuth of \mbox{\boldmath$k$}, in the response function $I(q,\omega)$ is performed over the interval $\phi\in\Phi(\mbox{\boldmath$q$},\omega)$ selected by equations
\bea
\varepsilon_+(\mbox{\boldmath$k$}+\mbox{\boldmath$q$}/2)\geq\mu,~
\varepsilon_-(\mbox{\boldmath$k$}-\mbox{\boldmath$q$}/2)\leq\mu,~\nonumber\\
\varepsilon_+(\mbox{\boldmath$k$}+\mbox{\boldmath$q$}/2)
-\varepsilon_-(\mbox{\boldmath$k$}-\mbox{\boldmath$q$}/2)=\hbar\omega.
\label{eq11}
\eea
The family of integration paths over $\phi$, parameterized by the transition frequency $\omega$, fills the integration area in the \mbox{\boldmath$k$}-plane. For $q, k_\alpha\ll k_F$, this area is close to a ring restricted by two circles with the radii $k_F\pm k_\alpha$, whose centers are displaced from $\mbox{\boldmath$k$}=0$ by $\pm\mbox{\boldmath$q$}/2$. The reactive part of spin polarization, phase shifted by $\pi/2$ with respected to the field, has the same order of magnitude but its analytical form is more involved.

The position of the band center, ${\bar\omega}=2v_F k_\alpha$, and the half-width of the band, $\omega_{1/2}=v_F(q+2k_\alpha^2/k_F)$, determine two characteristic scales in the frequency domain. For small $q$, $q\ll k_\alpha$, the band is narrow, $\omega_{1/2}\ll{\bar \omega}$, however, $\omega_{1/2}\rightarrow{\bar \omega}$ when $q\rightarrow 2k_\alpha$. The topology of the integration paths in the \mbox{\boldmath$k$}-plane changes at $q_{cr}=2k_\alpha^2/k_F$. For $q<q_{cr}$, the shape of the band evolves smoothly, Fig.~2. At $q=q_{cr}$, the band acquires a sharp peak, and for $q>q_{cr}$ it becomes wide and strongly asymmetric. When $q\rightarrow 2k_\alpha$, two circles in the \mbox{\boldmath$k$}-plane touch to one another, and the low-frequency wing of $I(q,\omega)$ moves to $\omega=0$ and diverges. Therefore, $2k_\alpha^2/k_F$ and $k_\alpha$ are two characteristic SO scales in the momentum domain. The divergence of $I(q,\omega)$ for $q\rightarrow 2k_\alpha$ underscores the importance of the electric field gradients for the production of nonequilibrium spins. For $q>2k_\alpha$, the collisionless regime is no more self-consistent for an infinite area.

Another contribution to $\sigma_x(\mbox{\boldmath$q$},\omega)$ comes from intraband transitions. When $q\ll k_F$, the in-phase magnetization does not vanish for $\omega<v_Fq$, i.e.,
\be
\sigma_x(\mbox{\boldmath$q$},\omega)\propto\Theta(v_Fq-\omega),
\label{eq11a}
\ee
where $\Theta(x)$ is a Heaviside function. Similar equations exist for spin currents ${\cal J}_{ij}$, while their specific form depends on the indeces $i,j$. Therefore, the intrabranch contributions to the magnetization and spin currents exist, should compete with the interbranch contributions, and are singular in the $q,\omega\rightarrow 0$ limit. This behavior is reminescent of the Schliemann and Loss discussion of the dependence of spin currents on the ratio of $\omega$ and $1/\tau_p$ \cite{SL04}. It is obvious from Eq.~(\ref{eq11a}) why the intrabranch contribution to spin magnetization disappears from the $\mbox{\boldmath$q$}=0$ responses in the clean limit.

 \centerline{\epsfig{file=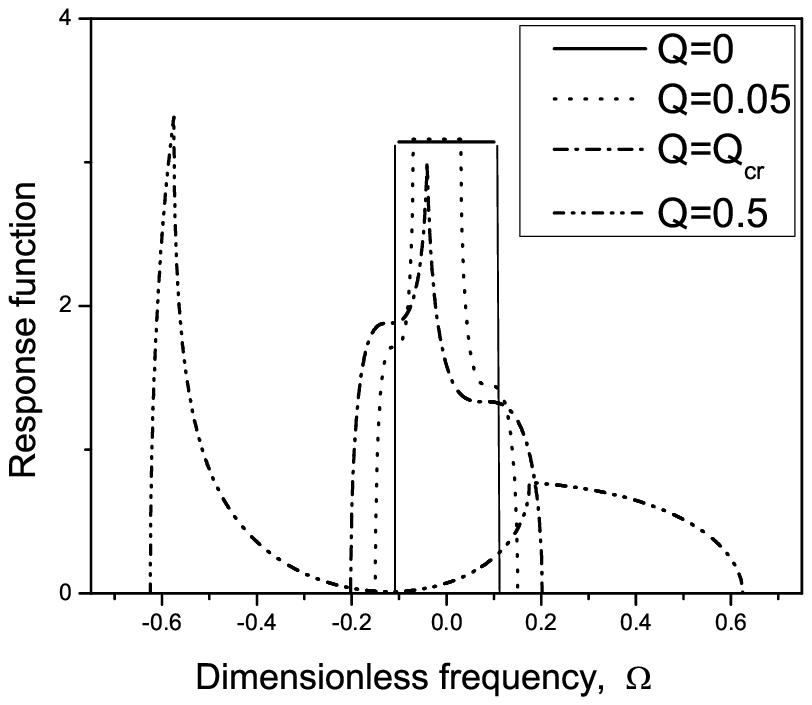,width=9cm}}
 \vskip 0.05 in
 \vbox{\baselineskip0.4 pt
 {\small {\bf Fig.~2} Response function $I(Q,\Omega)$ for the dynamic
spin magnetization {\it vs} dimensionless frequency
$\Omega=m(\omega-{\bar\omega})/2\hbar k_F k_\alpha$ for four values of
the dimensionless momentum $Q=q/2k_\alpha$; $k_\alpha/k_F=0.1$,
$Q_{cr}\approx 0.1$.}}

Eq.~(\ref{eq11a}) has been derived for the Hamiltonian ${\hat H}_R$. Its generalization for a $k$-dependent SO coupling constant, $\alpha_R=\alpha_R(k^2)$, should split the $\Theta$-function into two terms.

It is a remarkable fact that the characteristic length $k_\alpha^{-1}$ found here for the spin production coincides with the Dyakonov-Perel spin-relaxation length. Indeed, the spin relaxation time is $\tau_s^{-1}\approx(\alpha k_F)^2\tau_p/\hbar^2$, and therefore the spin diffusion length is $L_s\approx v_F(\tau_p\tau_s)\approx k_\alpha^{-1}$. This coincidence indicates strong entanglement of the production and relaxation processes in the inhomogeneous spin transport. Moreover, when $q\rightarrow 2k_\alpha$ and the low-frequency wing of the interbranch absorption softens, $({\bar\omega}-\omega_{1/2})\rightarrow0$, the inter- and intrabranch absorption merge.  

\subsection{Discussion and Conclusions}

The complexity of the spin transport in media with spin-orbit interaction produced controversies in the results of different theoretical groups that have not been resolved yet \cite{Mur,Sino,Mur2,R03,Burkov,XX04,Loss04,R04,Inoue,Olga,MSH,Nomura,Chulaev}. The unsettled problems include some analytical results, their physical interpretation, and the comparison of the analytical and numerical results. These problems seem to stem from the competition of different terms that is similar to the cancelations discussed in Sec.~III.B. However, the proper treatment of such a competition for nonequilibrium systems is much more demanding. 

In this paper, relation between spin currents and the SO contribution to the spontaneous electric polarization, as well as their connection to the dielectric function of noncentrosymmetric crystals, have been established. The effect of the crystal symmetry on the background (equilibrium) spin currents has been investigated, and the implications for spin currents driven by external electric fields have been discussed. Spin-response to an inhomogeneous time-dependent electric field has been found. 

As distinct from spin currents that have no rigorous theoretical justification and manifest itself in dielectric polarization, spin magnetization is a well defined quantity. From this standpoint, spin magnetization and electric polarization (electric current) can be considered as basic variables. They are coupled through spin-orbit interaction and their propagation is a collective effect \cite{Burkov,MSH}. Such an approach has the advantage of operating in terms of measurables, i.e., the quantities that can be controlled experimentally, and well matches Maxwellian electrodynamics. 

In the practical aspect, an electric field \mbox{\boldmath$E$} can produce spin magnetization only by breaking the time-inversion symmetry. When $\mbox{\boldmath$B$}=0$, it is possible either due to dissipation or by employing transients. The latter approach seem to have the advantage of lesser losses. Similarly, small structure sizes or large gradients are needed for a low-losses ballistic spin transport. Therefore, the geometries including transients and sharp gradients facilitate spin magnetization and ballistic spin fluxes. 

{\bf ACKNOWLEDGMENTS }

Funding of this research granted through a DARPA contract and the collaboration with Dr. Al. L. Efros (NRL) on the results presented in Sec.~II are gratefully acknowledged.


\begin{thebibliography}{99}
\bibitem[*]{Rashba*} Email: erashba@mailaps.org
\bibitem{Wolf} S. A. Wolf, D. D. Awschalom, R. A. Buhrman, J. M. Daughton, S. von Molnar, M. L. Roukes, A. Y. Chtchelkanova, and D. M. Treger, {\it Science} {\bf 294}, 1488 (2001).
\bibitem{Gregg} J. F. Gregg, I. Petej, E. Jouguelet, and C Dennis, {\it J. Phys. D: Appl. Phys.} {\bf 35}, R121 (2002).
\bibitem{Zutic} I. \v{Z}uti\'{c}, J. Fabian, and S. Das Sarma, {\it Rev. Mod. Phys.} {\bf 76}, 323 (2004).
\bibitem{ALS} D. D. Awschalom, D. Loss, and N. Samarth, {\it Semiconductor Spintronics and Quantum Computation} (Springer, Berlin, 2002).
\bibitem{R60} E. I. Rashba, {\it Sov. Phys. - Solid State} {\bf 2}, 1109 (1960).
\bibitem{RS91} E. I. Rashba and V. I. Sheka, in {\it Landau Level Spectroscopy}, G. Landwehr and E. I. Rashba, eds. (North-Holland, Amsterdam, 1991), pp. 131 - 206.
\bibitem{MBK} B. D. McCombe, S. G. Bishop, and R. Kaplan, {\it Phys. Rev. Lett.} {\bf 18}, 748 (1967).
\bibitem{Dob83} M. Dobrowolska, Y. F. Chen, J. K. Furdyna, and S. Rodriguez, {\it Phys. Rev. Lett.} {\bf 51}, 134 (1983).
\bibitem{PR65} S. I. Pekar and E. I. Rashba, {\it Sov. Phys. - JETP} {\bf 20}, 1295 (1965).
\bibitem{KRS93} L. S. Khazan, Yu. B. Rubo, and V. I. Sheka, {\it Phys. Rev.} B {\bf 47}, 13180 (1993).
\bibitem{Kato} Y. Kato, R. C. Myers, D. C. Driscoll, A. C. Gossard, J. Levy, and D. D. Awschalom, {\it Science} {\bf 299}, 1201 (2003).
\bibitem{REprl} E. I. Rashba and Al. L. Efros, {\it Phys. Rev. Lett.} {\bf 91}, 126405 (2003).
\bibitem{REapl} E. I. Rashba and Al. L. Efros, {\it Appl. Phys. Lett.} {\bf 83}, 5295 (2003).
\bibitem{Marcus} S. K. Watson, R. M. Potok, C. M. Marcus, and V. Umansky, {\it Phys. Rev. Lett.} {\bf 91}, 258301 (2003).
\bibitem{Rokh} L. P. Rokhinson, Y. B. Lyanda-Geller, L. N. Pfeiffer, and K. W. West, cond-mat/0403645.
\bibitem{Mishch} E. G. Mishchenko and B. I. Halperin, {\it Phys. Rev.} B {\bf 68}, 045317 (2003).
\bibitem{Silsbee} R. H. Silsbee, {\it J. Phys.: Condens. Matter} {\bf 16}, R179 (2004).
\bibitem{Gov} M. Governale, F. Taddei, and R. Fazio, {\it Phys. Rev.} B {\bf 68}, 155324 (2003).
\bibitem{Malsh} A. G. Mal'shukov, C. S. Tang, C. S. Chu, and K. A. Chao, {\it Phys. Rev.} B {\bf 68}, 233307 (2004).
\bibitem{Mur} S. Murakami, N. Nagaosa, and S.-C. Zhang, {\it Science} {\bf 301}, 1348 (2003).
\bibitem{Sino} J. Sinova, D. Culcer, Q. Niu, N. A. Sinitsyn, T. Jungwirth, and A. H. MacDonald, {\it Phys. Rev. Lett.} {\bf 92}, 126603 (2004).
\bibitem{Winkler} R. Winkler, {\it Spin-Orbit Coupling Effects in Two-Dimensional Electron and Hole Systems} (Springer, Berlin, 2003).
\bibitem{WJMR} Z. Wilamowski, W. Jantsch, H. Malissa, and U. R\"{o}ssler, {\it Phys. Rev.} B {\bf 66}, 195315 (2002).
\bibitem{MR} V. I. Mel'nikov and E. I. Rashba, {\it Sov. Phys. JETP} {\bf 34}, 1353 (1972).
\bibitem{FlMech} L. D. Landau and E. M. Lifshitz, {\it Fluid Mechanics} (Butterworth-Heinemann, Oxford 1999), \S 6 and \S 7.
\bibitem{DD90} S. Datta and B. Das, {\it Appl. Phys. Lett.} {\bf 56}, 665 (1990).
\bibitem{Mur2} S. Murakami, N. Nagaosa, and S.-C. Zhang, {\it Phys. Rev.} B {\bf 69}, 235206 (2004).
\bibitem{R03} E. I. Rashba, {\it Phys. Rev.} B {\bf 68}, 241315(R) (2003).
\bibitem{Burkov} A. A. Burkov, A. S. N\'{u}\H{n}es, and A. H. MacDonald, cond-mat/0311328.
\bibitem{XX04} Y. Xiong and X. C. Xie, cond-mat/0403083.
\bibitem{Loss04} S. I. Erlingsson, J. Schliemann, and D. Loss, cond-mat/0406531.
\bibitem{LP81} E. M. Lifshitz and L. P. Pitaevskii, {\it Physical Kinetics} (Pergamon, New York, 1981), \S 2.
\bibitem{ZawPf} W. Zawadzki and P. Pfeffer, {\it Semicond. Sci. Tech.} {\bf 19}, R1 - R17 (2003).
\bibitem{R04} E. I. Rashba, cond-mat/0404723.
\bibitem{Pareek} T. P. Pareek, {\it Phys. Rev. Lett.} {\bf 92}, 076601 (2004).
\bibitem{ML} F. Meier and D. Loss, {\it Phys. Rev. Lett.} {\bf 90}, 167204 (2003).
\bibitem{SKK} F. Sch\"{u}tz, P. Kopietz, and M. Kollar, cond-mat/0405312.
\bibitem{Stevens} M. J. Stevens, A. L. Smirl, R. D. R. Bhat, A. Najmaie, J. E. Sipe, and H. M. van Driel, {\it Phys. Rev. Lett.} {\bf 90}, 136603 (2003).
\bibitem{Dzy} I. E. Dzyaloshinskii, {\it Sov. Phys. JETP} {\bf 10}, 628 (1959).
\bibitem{IvchP} E. L. Ivchenko and G. E. Pikus, {\it JETP Lett.} {\bf 27}, 604 (1978).
\bibitem{Levit} L. S. Levitov, Yu. N. Nazarov, and G. M. Eliashberg, {\it Sov. Phys. JETP} {\bf 61}, 133 (1985).
\bibitem{Edel} V. M. Edelstein, {\it Solid State Commun.} {\bf 73}, 233 (1990).
\bibitem{Sharvin} Yu. V. Sharvin, {\it Sov. Phys. JETP} {\bf 21}, 655 (1965).
\bibitem{Vosk} A. Voskoboynikov, S. S. Liu, and C. P. Lee, {\it Phys. Rev.} B {\bf 59}, 12514 (1999).
\bibitem{Koga} T. Koga, J. Nitta, H. Takayanagi, and S. Datta,  {\it Phys. Rev. Lett.} {\bf 88}, 126601 (2002).
\bibitem{DP} M. I. D'yakonov and V. I. Perel', {\it Sov. Phys. - Solid State} {\bf 13}, 3023 (1972).
\bibitem{Aw04} Y. Kato, R. C. Myers, A. C. Gossard, and D. D. Awschalom, cond-mat/0403407. 
\bibitem{Gan04} S. D. Ganichev, S. N. Danilov, P. Schneider, V. V. Bel'kov, L. E. Golub, W. Wegscheider, D. Weiss, and W. Prettl, cond-mat/0403641.
\bibitem{SL04} J. Schliemann and D. Loss, {\it Phys. Rev.} B {\bf 69}, 165315 (2004).
\bibitem{Inoue} J. Inoue, G. E. W. Bauer, and L. W. Molenkamp, {\it Phys. Rev.} B {\bf 70}, 041303(R) (2004).
\bibitem{Olga} O. V. Dimitrova, cond-mat/0405339 and cond-mat/0407612.
\bibitem{MSH} E. G. Mishchenko, A. V. Shytov, and B. I. Halperin, cond-mat/0406730.
\bibitem{Nomura} K. Nomura, J. Sinova, T. Jungwirth, Q. Niu, and A. H. MacDonald, cond-mat/0407279.
\bibitem{Chulaev} O. Chulaev and D. Loss, cond-mat/0407342.
 

\end{thebibliography}
\end{document}